\documentclass[aps,prl,twocolumn,superscriptaddress]{revtex4-1}

\usepackage{graphicx}
\usepackage{color}
\usepackage{braket}
\usepackage{hyperref}
\usepackage{amssymb}

\usepackage{placeins}

\begin{document}

%\preprint{}

\title{Sound propagation and quantum limited damping in a two-dimensional Fermi gas}


\author{Markus Bohlen}

\affiliation{Institut f\"{u}r Laserphysik, Universit\"{a}t Hamburg}
\affiliation{The Hamburg Centre for Ultrafast Imaging, Universit\"{a}t Hamburg, Luruper Chaussee 149, 22761 Hamburg}
\affiliation{Laboratoire Kastler Brossel, ENS-Universit\'{e} PSL, CNRS, Sorbonne Universit\'{e}, Coll\`{e}ge de France, 24 rue Lhomond, 75005 Paris, France}

\author{Lennart Sobirey}
\author{Niclas Luick}
\affiliation{Institut f\"{u}r Laserphysik, Universit\"{a}t Hamburg}

\author{Hauke Biss}
\affiliation{Institut f\"{u}r Laserphysik, Universit\"{a}t Hamburg}
\affiliation{The Hamburg Centre for Ultrafast Imaging, Universit\"{a}t Hamburg, Luruper Chaussee 149, 22761 Hamburg}

\author{Tilman Enss}
\affiliation{Intitut f\"{u}r Theoretische Physik, Universit\"{a}t Heidelberg, Philosophenweg 19, 69120 Heidelberg}

\author{Thomas Lompe}
\email{tlompe@physik.uni-hamburg.de}
\affiliation{Institut f\"{u}r Laserphysik, Universit\"{a}t Hamburg}

\author{Henning Moritz}
\affiliation{Institut f\"{u}r Laserphysik, Universit\"{a}t Hamburg}
\affiliation{The Hamburg Centre for Ultrafast Imaging, Universit\"{a}t Hamburg, Luruper Chaussee 149, 22761 Hamburg}

%\date{\today}

\begin{abstract}
%2D-systems are fascinating because the reduced dimensionality does crazy shit.
Strongly interacting two-dimensional Fermi systems are one of the great remaining challenges in many-body physics due to the interplay of strong local correlations and enhanced long-range fluctuations.
Here, we probe the thermodynamic and transport properties of a 2D Fermi gas across the BEC-BCS crossover by studying the propagation and damping of sound modes.
We excite particle currents by imprinting a phase step onto homogeneous Fermi gases trapped in a box potential and extract the speed of sound from the frequency of the resulting density oscillations.
We measure the speed of sound across the BEC-BCS crossover and compare the resulting dynamic measurement of the equation of state both to a static measurement based on recording density profiles and to Quantum Monte Carlo calculations and find reasonable agreement between all three.
We also measure the damping of the sound mode, which is determined by the shear and bulk viscosities as well as the thermal conductivity of the gas. 
We find that the damping is minimal in the strongly interacting regime and the diffusivity approaches the universal quantum bound $\hbar/m$ of a perfect fluid.
\end{abstract}

%\keywords{}

\maketitle

Strongly interacting fermionic systems appear in many different areas of physics, yet understanding their behavior remains challenging. 
A powerful experimental method to gain access to their thermodynamic and transport properties is to study collective excitations such as sound modes. 
The speed of sound is determined by the compressibility of the medium, giving access to its equation of state.
The damping of sound modes in the hydrodynamic regime is caused by the diffusion of heat as well as longitudinal and transverse momentum, and thus depends on the transport coefficients of the medium, i.e., thermal conductivity, bulk and shear viscosity. 
In hydrodynamic systems which support well-defined quasiparticle excitations such as fermionic quasiparticles in a normal Fermi liquid or phonon excitations in a superfluid, kinetic theory predicts that the damping rate is proportional to the quasiparticle lifetimes. 
Long-lived quasiparticles can transport heat or momentum over long distances and therefore smooth out density and pressure variations very efficiently, leading to strong attenuation of sound waves.
In contrast, no well-defined quasiparticles exist in the strongly correlated regime.
Here, particles scatter with a mean free path comparable to the average interparticle spacing, leading to lower diffusivities and hence lower damping rates.
A lower bound for the diffusivities $D\gtrsim \hbar/m$ and thus quantum limited transport has been predicted and observed in several transport channels, including shear viscosity in ultracold 2D \cite{Vogt2012,Bruun2012,Schafer2012,Enss2012a} and 3D \cite{Massignan2005,Enss2011,Cao2011} Fermi gases as well as spin diffusion in 2D \cite{Bruun2012,Enss2012a,Koschorreck2013,Luciuk2017} and 3D Fermi gases \cite{Sommer2011,Bruun2011,Enss2012b,Bardon2014,Valtolina2017}.
In measurements of sound propagation, this limit was observed in the sound diffusion of a unitary 3D Fermi gas \cite{Patel2019}.

Several hypotheses have been brought forward to provide an explanation for quantum limited transport \cite{Enss2019}:
One, motivated by holographic duality \cite{Kovtun2005}, is that it occurs near scale invariant points in the phase diagram.
The unitary 3D Fermi gas is an example that seems to support this hypothesis since it is strongly interacting as well as scale invariant and exhibits quantum limited shear and spin diffusion.
In contrast, 2D Fermi gases exhibit a quantum scale anomaly that breaks scale invariance \cite{Olshanii2010,Holten2018,Peppler2018,Murthy2019}. 
Here, we investigate the propagation and damping of sound in a strongly interacting 2D Fermi gas and thereby probe a crucial test case for this hypothesis.
We observe that the damping approaches the quantum limit $D\approx\hbar/m$ in the strongly interacting regime, where scale invariance is most dramatically violated, showing that scale invariance or quantum criticality is in fact not required for quantum limited transport.
Similar observations were made for transverse spin diffusion in \cite{Luciuk2017}.
Our results confirm a scenario of incoherent transport that has emerged in recent years from the study of anomalous transport in high-temperature superconductors and other 'bad metals' \cite{Bruin2013,Hartnoll2015,Hartnoll2018} and links quantum limited transport to strong correlations.

\begin{figure}
	\center
	\includegraphics[width=8.6cm]{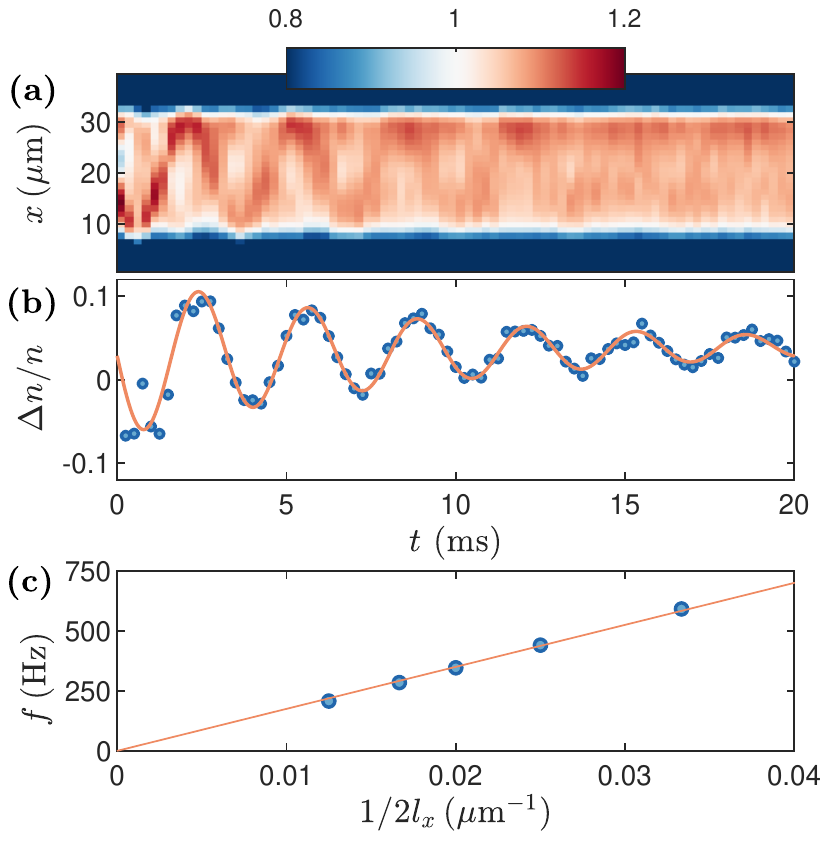}
	\caption{\label{Fig. 1}
	Propagation of a sound wave in a box potential: 
	(a) Density profiles $n(x,t)/n(t)$ averaged along the direction perpendicular to the sound propagation and normalized to the average density $n(t)$ within the box.
	A density wave propagating through the box is visible.
    Each profile $n(x,t)/n(t)$ is the average of 120 individual realizations.
	Note that the color scale has been chosen to enhance the visibility of the sound wave. 
	(b) Relative density imbalance between the two sides of the box for the same data set.
	The solid line shows a damped sinusoidal fit to the data.
	(c) Frequency of the density oscillation as a function of the inverse box length.
	The slope of the linear fit (solid line) corresponds to the speed of sound.
	In (b) and (c), the statistical errors are smaller than the marker size.
	Each data point in (c) is the average of 23 realizations.
	}
\end{figure}
We perform our studies of sound propagation with an ultracold gas of $^6$Li atoms in a spin-balanced mixture of the lowest two hyperfine states, trapped in a two-dimensional box potential \cite{Hueck2018,Luick2019}.
The gas is vertically confined in a single node of a repulsive optical lattice with trap frequency $\omega_z/2\pi = 8.4(3)\,\mathrm{kHz}$.
At our densities of about $n_{\uparrow/\downarrow} \equiv n \approx 1\,\mathrm{\mu m}^{-2}$ per spin state, the chemical potential $\mu$ is smaller than the vertical level spacing $\hbar \omega_z$, which ensures that our gas is in the quasi 2D regime.
The confinement in the horizontal plane is created using a digital micromirror device illuminated with blue detuned light ($\lambda =532\,$nm), trapping the gas in a two-dimensional box with a typical size of $l_x \times l_y = 30 \times 40 \,\mathrm{\mu m}^2$.
According to the temperature determination performed in \cite{Luick2019}, our system is in the low-temperature regime with $k_\mathrm{B} T/E_\mathrm{F} \leq 0.1$, where $E_\mathrm{F} = \hbar^2 k^2_\mathrm{F}/2m$ is the Fermi energy, $k_\mathrm{F} = (4\pi n)^{1/2}$ the Fermi momentum and $m$ the atomic mass of $^6$Li.

For our experiments, we build on the experimental procedure developed in \cite{Ville2018}, where sound propagation was studied in weakly interacting 2D Bose gases.
To excite a sound mode in the box we follow the approach of \cite{Luick2019} and imprint a relative phase between two halves of the system by illuminating one side with a spatially homogeneous optical potential for a short duration $\tau < h/E_{\text F}$.
We then observe the resulting density oscillations by imaging the density distribution after different hold times using in-situ absorption imaging.
An example of such an oscillation is shown in figure 1(a).
A sound wave traveling back and forth between the two sides of the box is clearly visible in the density profile.
To extract the oscillation frequency $f=\omega/2 \pi$ and the damping $\Gamma$ of this sound wave, we calculate the relative particle imbalance $\Delta n/n = 2 (n_\mathrm{t} - n_\mathrm{b}) / (n_\mathrm{t}+n_\mathrm{b})$ from the densities $n_\mathrm{t}$ and $n_\mathrm{b}$ in the top and bottom halves of the box and fit it with a damped sinusoidal of the form $A(t) = A_0 \cos{(\omega t + \phi)} \exp{(-\Gamma t/2)} + b$ (see figure 1(b)).
We measure the oscillation frequency for different boxes with lengths between $l_x = 15\,\mathrm{\mu m}$ and $l_x = 40\,\mathrm{\mu m}$ and find that it is proportional to the inverse of the box length (see figure 1(c)). 
This confirms that we observe a sound wave traveling at a constant velocity $c = 2 l_x f$ and that edge effects are negligible.

\begin{figure}
	\center
	\includegraphics[width=8.6cm]{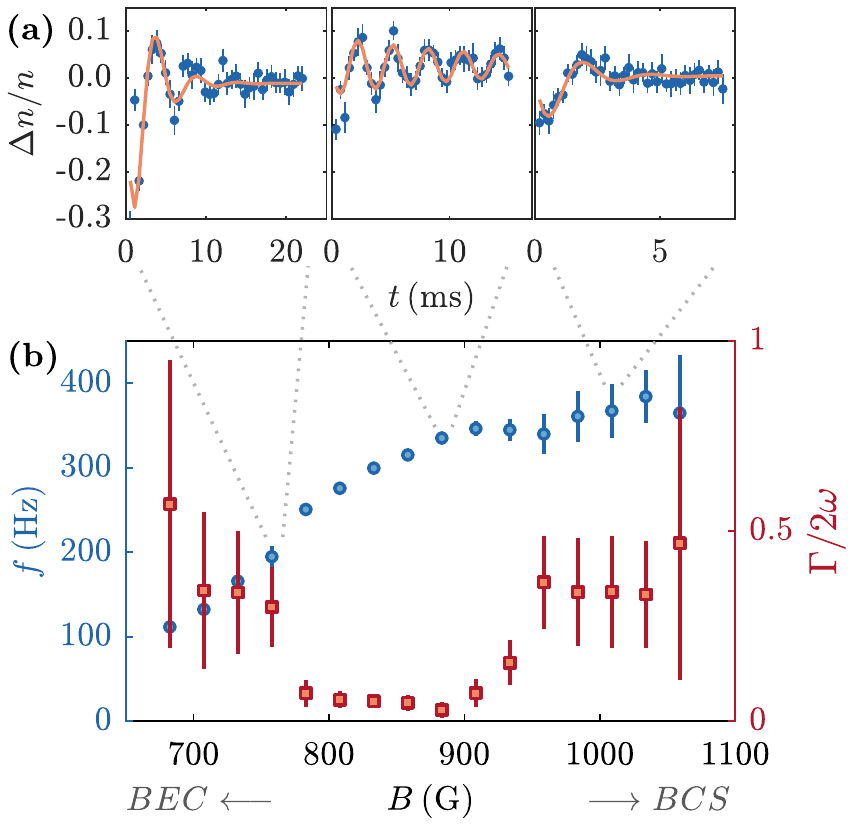}
	\caption{\label{Fig. 2} 
	(a): Oscillations in the density imbalance at $B=758\,\mathrm{G}$, $B=883\,\mathrm{G}$ and $B=1009\,\mathrm{G}$. 
	Solid lines represent damped sinusoidal fits. 
	(b): Frequency (blue circles) and damping (red squares) of sound oscillations as a function of magnetic field.
	The frequency increases smoothly from the BEC to the BCS side and starts to saturate at high magnetic fields.
	The damping shows a clear minimum in the strongly interacting regime and increases strongly towards either side.
	Each data point is the average of 39 realizations.}
\end{figure}

\begin{figure*}
	\center
	\includegraphics[width=17.2cm]{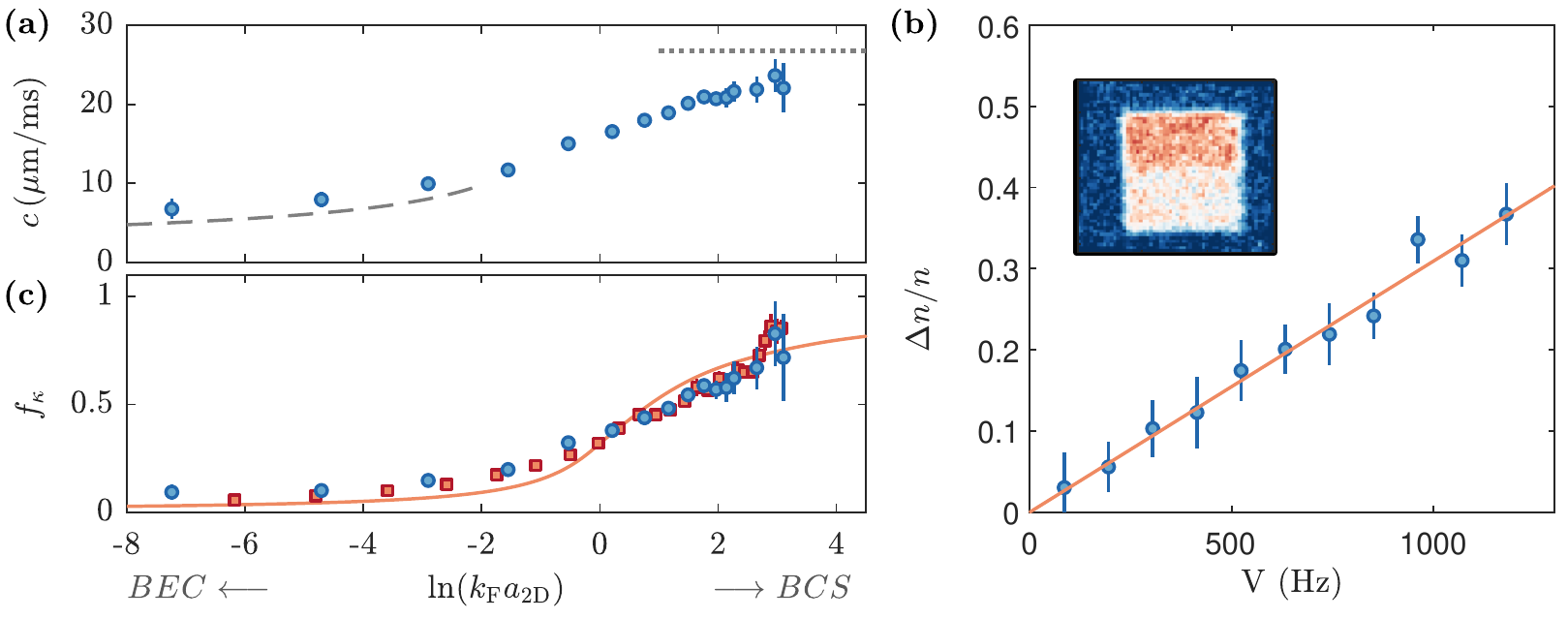}
	\caption{\label{Fig. 3} 
	(a): Speed of sound across the BEC-BCS crossover. 
	Gray lines represent the two theoretical limits: 
	In the BEC regime, Bogoliubov theory predicts $\mu = gn$ and hence $c_\mathrm{B} = (gn/2m)^{1/2}$, where $g$ is the interaction parameter between bound dimers defined as in \cite{Turlapov2017,Luick2019}.
	On the BCS side, $\mu \rightarrow E_\mathrm{F}$ yields a constant sound velocity $c_\mathrm{F} = v_\mathrm{F}/\sqrt{2}$ where $v_\mathrm{F} = \hbar k_\mathrm{F}/m$ is the Fermi velocity.
	(b): Static measurement of the compressibility. 
	A repulsive potential $V$ is imprinted onto one half of the box resulting in a density imbalance $\Delta n/n$ (blue points).
	We extract the compressibility from the initial slope of the data points according to $1/n^2\kappa = \lim_{\Delta n\rightarrow0}{V/\Delta n}$ in a local density approximation.
	Each data point and the inset are averages of 20 realizations. 
	(c): Comparison between the compressibility scaling functions $f_\kappa$ obtained from the speed of sound (blue circles), the density response to an imprinted static potential (red squares) and QMC calculations \cite{Shi2015,footnote_fkappa}.
	} 
\end{figure*}

To probe the speed of sound as a function of interaction strength, we perform measurements in a box with $l_x = 30\,\mathrm{\mu m}$ at magnetic fields around the Feshbach resonance.
Examples of the resulting oscillations as well as the evolution of the oscillation frequency as a function of magnetic field are shown in figure 2.
As the field is varied from the BEC side to the BCS side of the crossover, the oscillation frequency increases, which is expected since the compressibility of a Fermi gas is much lower than that of a weakly repulsive Bose gas.
On the Fermi side, the gas is thus stiffer with respect to density fluctuations and sound waves propagate faster than on the Bose side.

We plot the speed of sound extracted from the oscillation frequencies as a function of the 2D interaction parameter $\ln{(k_\mathrm{F}a_\mathrm{2D})}$ 
\footnote{
The 2D scattering length is defined as $a_\mathrm{2D} = a_\mathrm{2D}^{(0)} \exp(-0.5 \Delta w(\mu/\hbar \omega_z))$.
Here, $a_\mathrm{2D}^{(0)} = l_z \sqrt{\pi/0.905} \exp (-\sqrt{\pi/2}\, l_z/a_\mathrm{3D})$ is the 2D scattering length for dilute gases, with the 3D-scattering length $a_\mathrm{3D}$, the harmonic oscillator length in the strongly confined direction $l_z=\sqrt{\hbar/m\omega_z}$ and a positive scaling function $\Delta w(\mu/\hbar\omega_z) \lesssim 2.3$ which accounts for the energy dependence of 2D scattering \cite{Petrov2001,Boettcher2016}.
}
in figure 3(a).
In a superfluid gas, two-fluid hydrodynamics predict the occurrence of two sound modes which propagate at different velocities, as observed in \cite{Sidorenkov2013}.
These modes generally mix density and entropy degrees of freedom.
For strongly interacting 2D-superfluids however, density and entropy excitations have been predicted to be well decoupled \cite{Hu2014,Ota2018}, and hence the sound mode we observe should correspond to an almost pure density wave.
In this case, the velocity of a sound wave is given by 
\begin{equation}
c = \sqrt {\frac{n}{m}\frac{\partial \mu}{\partial n}\bigg|_s}
\label{eq:csound}
\end{equation}
and is direcly related to the isentropic compressibility $\kappa = \frac{1}{n^2}\frac{\partial n}{\partial \mu}\big|_s$  
\footnote{
We use the adiabatic compressibility since the sound frequency is large compared to the damping ($\omega \geq \Gamma$). 
}.
This relation gives us simple zero-temperature expressions for the speed of sound in the BEC and BCS limits of the crossover, which are in good agreement with our data (dashed and dotted lines in figure 3(a)).

For a quantitative analysis we use Eq.\ (\ref{eq:csound}) to extract the compressibility of our gas from our measurement of the speed of sound.
From this, we then determine the dimensionless inverse compressibility scaling function $f_\kappa = 1/nE_\mathrm{F} \kappa$ of a two-dimensional Fermi gas.

In addition to this dynamic measurement of the equation of state (EOS), we also perform a static measurement of the compressibility EOS by determining the density response of our system to a static repulsive potential, similar to the work performed in \cite{Makhalov2014,Boettcher2016,Fenech2016,Hueck2018} (see figure 3(b)).
This results in two independent measurements of $f_\kappa$, which show good agreement with each other (see figure 3(c)).

Finally we compare our data to theory by extracting $f_\kappa$ from Quantum Monte Carlo (QMC) calculations of the ground state energy $E_0$ of a homogeneous 2D Fermi gas \cite{Shi2015,footnote_fkappa}.
On the BCS side, the experimental results agree well with the theoretical prediction.
On the BEC side, both the static and the dynamic measurements lie above the theoretical prediction.
One possible explanation for this difference could be finite temperature effects, which have been observed in \cite{Boettcher2016} to decrease the compressibility of the gas, thus increasing $f_\kappa$.

Very recent simulations of the sound velocity in a 2D Bose gas \cite{Singh2020} indicate that the density and entropy modes remain coupled even for relatively strongly interacting Bose gases, leading to sound velocities which differ from the Bogoliubov prediction.
This should be observable as a difference between the static and dynamic measurements of the compressibility.
However, if such a difference exists in our system, it is smaller than the uncertainty of our measurement.

We now turn our attention to the damping of the sound waves.
In our strongly correlated system, the mean free path $l_\mathrm{mfp}$ of the particles is much smaller than the oscillation wavelength and their collision rate is high with respect to the oscillation frequency 
\footnote{
Within kinetic theory, the mean free path is estimated as $l_\mathrm{mfp} = 1/n\sigma \simeq 0.63 n^{-1/2}(1+4\ln^2(k_\mathrm{F} a_\mathrm{2D})/\pi^2)$. 
For our system, one obtains $l_\mathrm{mfp} \leq 16\,\mathrm{\mu m} < 2 l_x$.
Hydrodynamic estimates for the relaxation length $l=D_s/c$ and relaxation time $\tau=l/c$ yield $\Gamma/\omega \sim \omega\tau \ll 1$. 
Thus, both of these estimates suggest that hydrodynamic conditions are satisfied for our system.
}.
Hence the system is in the hydrodynamic regime.
In this regime, the spatial variations in density and temperature that constitute a sound wave lead to diffusive currents of longitudinal momentum, transverse momentum and heat, whose magnitudes are proportional to the bulk and shear viscosities $\zeta$ and $\eta$ and to the heat conductivity $\beta$ \cite{LandauLifshitzHydro,Smith1989}.
These diffusive currents smooth out the density and temperature variations and thus damp the sound wave according to the sound diffusion constant
\begin{equation}
D_s = \frac{\eta}{\rho} + \frac{\zeta}{\rho} + \beta \frac{c_p-c_v}{c_p c_v} = \Gamma/k_0^2 ,
\end{equation}
where $k_0 = \pi/l_x$ is the wave vector of the sound wave and $c_p$ and $c_v$ are the heat capacities at constant pressure and volume. 
\begin{figure}
	\center
	\includegraphics[width=8.6cm]{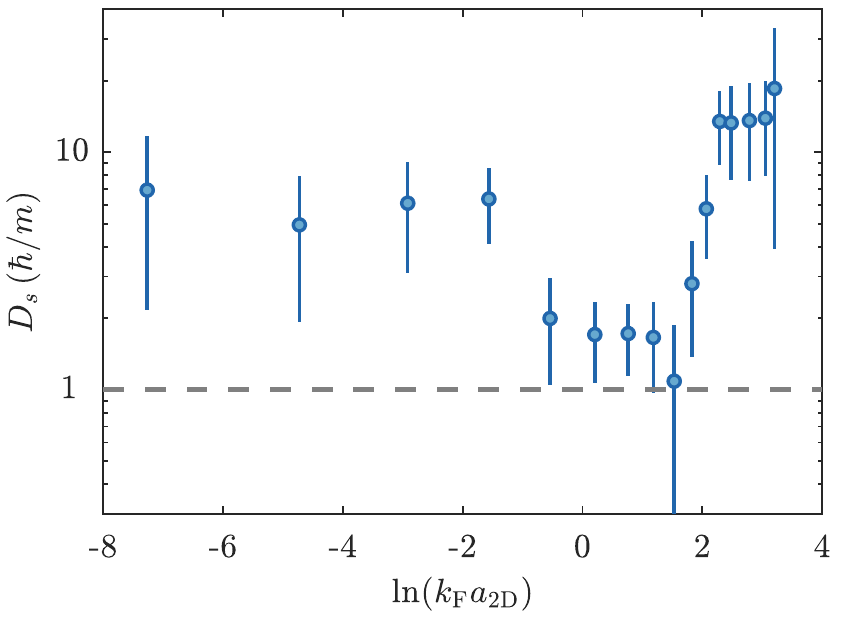}
	\caption{\label{Fig. 4} 
	Sound diffusion coefficient across the BEC-BCS crossover.
	In the strongly correlated regime, the diffusion coefficient reaches a minimum which agrees well with the universal quantum bound for diffusion at $\hbar/m$ (dashed line).
	}
\end{figure}
The evolution of $D_s$ across the BEC-BCS crossover is shown in figure 4.
It exhibits a broad minimum in the crossover regime $-1< \ln{(k_\mathrm{F}a_\mathrm{2D})}<2$ and increases steeply towards the BEC and BCS limits.

Before comparing our data to theory, we note that making quantitative predictions for transport coefficients of strongly interacting 2D Fermi gases in the low-temperature regime is still a major theoretical challenge.
%Regarding the comparison of our data to theory, we first note that quantitative predictions for transport coefficients of strongly interacting 2D Fermi gases in the low-temperature regime are not available to our best knowledge.
%We now compare our data to theory.
%We are not aware of quantitative predictions for transport coefficients of strongly interacting 2D Fermi gases in the low-temperature regime that we can compare out data to.
%Unfortunately, comparing our data to theory is difficult, as we are not aware of quantitative predictions for transport coefficients of strongly interacting 2D Fermi gases in the low-temperature regime.
Approaches such as Fermi liquid theory and BCS theory \cite{Smith1989} are only accurate at weak coupling.
Results obtained for the high-temperature regime ($T \geq T_\mathrm{F}$) indicate that the shear viscosity and heat conductivity have a minimum in the strongly correlated regime \cite{Enss2012a,Schafer2012,Bruun2012} whereas the bulk viscosity is maximal near resonance yet contributes much less to the damping \cite{Enss2019a,Nishida2019,Hofmann2020a}.
In total, high-temperature theory predicts a minimum of the sound diffusion in the crossover regime.
Although our measurements are performed in the low-temperature regime, the observed behavior is in qualitative agreement with an extrapolation of the high-temperature result to the low-temperature regime.

A prediction for a lower bound of $D_s$ in the strongly interacting regime can be obtained via a simple scaling argument:
In kinetic theory, the diffusion coefficient is given by the mean free path and the velocity via $D_s\sim v\, l_\mathrm{mfp}$. 
For a strongly interacting degenerate gas the mean free path $l_\mathrm{mfp}$ is on the order of the interparticle separation $n^{-1/2}$ and the velocity on the order of the Fermi velocity $v\sim v_\mathrm{F}\sim\hbar n^{1/2}/m$, resulting in a diffusion coefficient $D_s\sim  \hbar/m$. 
Since the interparticle separation is a lower limit for the mean free path, this yields a generic lower bound for the damping.
This lower limit is in agreement with our measured diffusion coefficient of $D_s\approx 1.8(2)\hbar/m$ in the strongly correlated regime.
Thus our strongly interacting 2D Fermi gas is a nearly perfect fluid despite the fact that scale invariance is broken and that the system is not at a quantum critical point.

In this work, we have studied the propagation and damping of sound waves in a homogeneous 2D Fermi gas across the BEC-BCS crossover.
We have extracted the compressibility EOS from a measurement of the speed of sound and find good agreement with $T=0$ QMC calculations.
We have measured the sound diffusion constant as a function of the interaction strength and find universal sound diffusion $D_s \sim \hbar/m$ and quantum limited transport in the strongly interacting regime.
This lower limit is reached at interactions where scale invariance is violated most severely \cite{Luciuk2017, Murthy2019}, but where the mean free path is comparable to the particle spacing. 
Since sound diffusion is the sum of momentum and heat diffusion, we thus find upper bounds of order $\hbar/m$ on each diffusion channel individually in the crossover regime. 
This demonstrates that the 2D Fermi gas realizes a nearly perfect fluid \cite{Kovtun2005, Schafer2009} and provides a benchmark against which future theoretical predictions can be validated.

An interesting extension of our measurements would be to study the temperature dependence of $D_s$ as done in unitary 3D Fermi gases \cite{Patel2019}.
In the fermionic regime, this would allow us to observe whether there is a maximum of $D_s$ at the critical temperature of superfluidity, similar to measurements in $^3$He \cite{Eska1980}.
In the deep BEC regime, control over the temperature of the gas would enable studies of second sound in a strongly interacting Bose gas \cite{Ota2018,Singh2020}.

\begin{acknowledgments}
We thank L. Mathey, V. Singh, A. Sinatra, Y. Castin and N. Defenu for stimulating discussions.
We also thank K. Morgener, K. Hueck and W. Weimer for their work in the early stages of the experiment.
This work is supported by the European Union's Seventh Framework Programme (FP7/2007-2013) under grant agreement No. 335431 and by the Deutsche Forschungsgemeinschaft (DFG, German Research Foundation) in the framework of SFB 925 and "SFB 1225" (ISOQUANT), the excellence cluster 'Advanced Imaging of Matter' - EXC 2056 - project ID 390715994 and under Germany’s Excellence Strategy “EXC-2181/1- 390900948” (the Heidelberg STRUCTURES Excellence Cluster).
M. Bohlen acknowledges support by Labex ICFP of \'{E}cole Normale Sup\'{e}rieure Paris.
\end{acknowledgments}

\bibliography{Sound}

\end{document}